\documentclass[
reprint,
showpacs,preprintnumbers,
amsmath,amssymb,
aps,
pra
]{revtex4-1}

\usepackage{graphicx}
\usepackage{dcolumn}
\usepackage{bm}
\usepackage{amsmath}
\usepackage{graphicx}
\usepackage{amssymb}
\usepackage{braket}
\usepackage{colortbl} 
\usepackage[usenames,dvipsnames,svgnames,table]{xcolor}
\usepackage[breaklinks]{hyperref}
\hypersetup{        
    unicode=false,     
    linktocpage=true,     
    pdftoolbar=true,      
    pdfmenubar=true,        
    pdffitwindow=false,     
    pdfstartview={FitH},     
    colorlinks=true,       
    linkcolor=WildStrawberry,   
    citecolor=CornflowerBlue,   
    filecolor=BrickRed,      
    urlcolor=Blue}
\usepackage{natbib}

\begin{document}

\preprint{APS/123-QED}

\title{
Experiment to detect dark energy forces using atom interferometry
}

\author{D. O. Sabulsky}
\altaffiliation{Now at Laboratoire Photonique, Num\'{e}rique et Nanosciences,\\ Universit\'{e} Bordeaux-IOGS-CNRS:UMR 5298, \\rue F. Mitterrand, F-33400 Talence, France}
\author{I. Dutta}
\affiliation{Centre for Cold Matter, Blackett Laboratory, Imperial College London, Prince Consort Road, London SW7 2AZ, United Kingdom}
\author{E. A. Hinds}
\email{ed.hinds@imperial.ac.uk}
\affiliation{Centre for Cold Matter, Blackett Laboratory, Imperial College London, Prince Consort Road, London SW7 2AZ, United Kingdom}

\author{B. Elder}
\affiliation{School of Physics and Astronomy, University of Nottingham, Nottingham, NG7 2RD, United Kingdom}
\author{C. Burrage}
\affiliation{School of Physics and Astronomy, University of Nottingham, Nottingham, NG7 2RD, United Kingdom}
\author{Edmund J. Copeland}
\affiliation{School of Physics and Astronomy, University of Nottingham, Nottingham, NG7 2RD, United Kingdom}

\date{\today} 

\begin{abstract}

The accelerated expansion of the universe motivates a wide class of scalar field theories that modify gravity on large scales. 
In regions where the weak field limit of General Relativity has been confirmed by experiment, such theories need a screening mechanism to suppress the new force. We have measured the acceleration of an atom toward a macroscopic test mass inside a high vacuum chamber, where the new force is unscreened in some theories. Our measurement, made using atom interferometry, shows that the attraction between atoms and the test mass does not differ appreciably from Newtonian gravity. This result places stringent limits on the free parameters in chameleon and symmetron theories of modified gravity.

\end{abstract}

\pacs{enter here}
\maketitle

The accelerating expansion of the universe and the uneven distribution of light and matter within it, have led to the conclusion that most of the energy in the universe is `dark' \cite{Copeland2006}. The nature of this energy is not yet understood, but it
motivates us to take seriously the notion that Einstein gravity may receive $O(1)$ corrections from new degrees of freedom.  However, measurements of the gravitational attraction between masses show that the force associated with such a field is far weaker than Newtonian gravity, suggesting that the force is either intrinsically very weak or some mechanism screens the force. Two suitably screened fields are the chameleon \cite{Khoury2004} and the symmetron \cite{Hinterbichler2010}, which couple both to themselves and to matter.  It is now understood \cite{Wang:2012kj} that these particular theories cannot drive accelerated expansion alone without some form of dark or vacuum energy, yet they remain interesting as a general way to hide modifications to Einstein gravity.

We proposed \cite{Burrage2015} that these theories could be tested very stringently by using atom interferometry to measure the force, in ultra-high vacuum, between a macroscopic test mass and an atom. Our proposal was pursued by Hamilton \textit{et al.} \cite{Hamilton:2015zga, Elder:2016yxm, Jaffe2017}. Here we report an independent atom interferometry measurement, using a significantly different experimental method, which achieves very similar sensitivity and confirms that a large part of the theoretical parameter space is now excluded. 

\begin{figure}
\centering
\includegraphics[width=.8\linewidth]{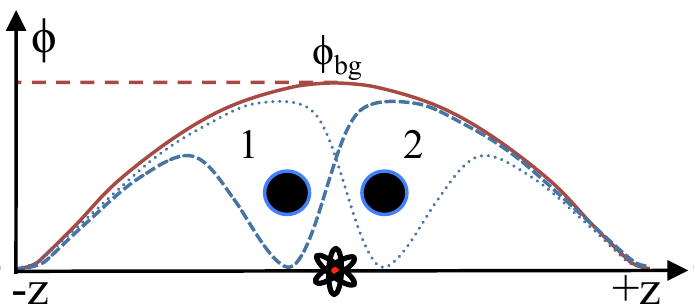}
\caption{
Principle of the experiment \cite{Burrage2015}. Vacuum chamber walls are at $\pm$Z. Solid red curve: scalar field $\phi$ is small at $\pm$Z, rising to $\phi_{\text{bg}}$ at the centre of the empty chamber. Dashed(dotted) blue curve:  ball in position 1(2) perturbs $\phi$ to produce a gradient $\nabla\phi$. Atoms at the centre of the chamber have acceleration $a_{\phi}\propto\nabla\phi$ toward the ball, which we measure by atom interferometry. \label{fig:principle}}
\end{figure}

Figure \ref{fig:principle} illustrates the principle of the experiment. The walls of the stainless steel vacuum chamber are at $\pm$Z, where the high density of the material forces the chameleon or symmetron field $\phi$ to a low value. Inside the empty vacuum chamber $\phi$ rises to a maximum, $\phi_{bg}$, illustrated by the red curve. A metal ball placed in the vacuum at position 1 forces a dip in $\phi_{bg}$ (dashed blue line), creating a gradient of $\phi$ in the vicinity of the ball. An atom placed at the centre of the chamber is accelerated by this gradient toward the ball, with acceleration $a_{\phi}$. We use  atom interferometery to measure the component of $a_{\phi}$ along the axis marked $\pm Z$. When the ball is moved to position 2 (dotted blue line), this component of $a_{\phi}$ reverses, and we detect that change of acceleration.

Atom interferometry with stimulated Raman transitions \cite{Kasevich1991,Peters1999} is a most sensitive technique for detecting small accelerations. It is already well established as the basis for ultra-sensitive gravimeters \cite{Landragin2010}, gyroscopes \cite{Kasevich1997}, magnetometers \cite{Narducci2014}, and accelerometers \cite{Bouyer2011} and for applications in metrology \cite{Weiss1993,Biraben2011}, tests of general relativity \cite{Kasevich2007}, and gravitational waves detection \cite{Kasevich2009,Bouyer2016,2018NatSR...814064C}. In our experiment, counter-propagating laser beams along the $z$ axis, differing in frequency by 6.8 GHz, drive the clock transition in rubidium-87 atoms through a Raman process. At time $t=0$ a $\pi/2$ pulse creates a superposition of the two clock states, which move apart because of the photon recoil momentum. After a time $T$, a $\pi$ pulse swaps the two internal states and reverses the recoil velocity so that the two parts of the wavefunction come back together at time $t=2T$. A final $\pi/2$ pulse then closes the interferometer to give $\cos^2(\varphi/2)$ and $\sin^2 (\varphi/2)$ fringes in the populations of the two clock states, where $\varphi$ is the quantum mechanical phase difference accumulated along the two paths. We determine $\varphi$ by reading out the final populations in the two clock states, and this phase is proportional to the acceleration of the atoms along the direction of the laser beams.  We look for the change in $\varphi$ when the ball is moved between positions 1 and 2 and use that to detect the gradient of the scalar field $\phi$.

\begin{figure}[t]
\centering 
\includegraphics[width=1\linewidth]{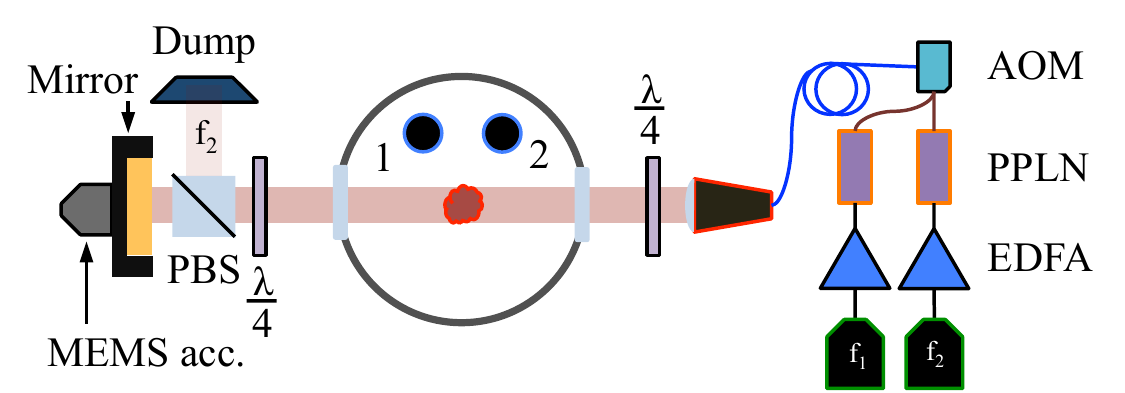}
\caption{
Lasers beams of frequency $f_1$ and $f_2$  are coupled into the two crossed linear polarisations of a polarisation-maintaining optical fibre. The fibre output is expanded and collimated to form a beam of 20.9\,mm radius ($e^{-2}$ intensity), which passes though a quarter-wave plate to make the two frequency components oppositely circularly polarised. A second quarter-wave plate restores the linear polarisation so that a polarising beam splitter can dump the $f_2$ beam, while a mirror retro-reflects the $f_1$ beam. At the atom cloud, the counter-propagating $f_1$ and $f_2$ beams have the same circular polarisation, as required to drive the $^{87}\text{Rb}$ clock transition. \label{fig:raman}}
\end{figure}

Inside the vacuum chamber, the density of residual gas is dominated by $9.6\times10^{-10}\,$mbar of H$_2$. The ball is a 19\,mm-radius sphere of aluminium, coated with Alion MH2200 paint to minimise the scattering of laser beams. This is suspended from a 6\,mm-thick aluminium rod, eccentrically mounted on a rotary vacuum feedthrough, which allows the position of the ball to be changed, as indicated in Fig.~\ref{fig:principle}. A 2D magneto-optical trap (MOT) \cite{Diekmann1998} injects a pulse of cold $^{87}$Rb atoms into the chamber through a differential pumping hole. A 3D MOT \cite{Wieman1992} collects $\sim 10^{8}$ atoms at the centre of the chamber, then optical molasses cools them to 5 $\mu$K before the cooling light is switched off, followed by the repump light once it has pumped all the atoms into the $5S_{1/2}(F=2)$ states. 

Next, we use Raman transitions to initiallise and operate the atom interferometer. The light is delivered to the atoms as shown in Fig. \ref{fig:raman}. Lasers of frequency $f_1$ and $f_2$ are phase locked to make a beat note, detuned by $\Delta f$ from the $6.8\,$GHz clock transition $\ket{F=2,M_F=0} \rightarrow \ket{F=1,M_F=0}$, which is resolved from the other hyperfine transitions by the Zeeman shifts in a $1.7\,$G magnetic field. The resonance frequency is Doppler shifted according to the atom's velocity component $v_{\parallel}$ along the Raman beams. With a detuning of $\Delta f=-72\,$kHz and a pulse length of $4.5\,\mu$s, atoms in the velocity range $v_{\parallel}=30\pm 23\,$mm/s are driven to the state $\ket{F=1,M_F=0}$, after which we blow away the remaining $F=2$ atoms using resonant light pressure.

This state-selected, velocity-selected group of $N_0 \simeq 10^6$ atoms is then subjected to the three interferometer pulses,  equally spaced by time intervals $T = 16\,$ms and having pulse areas of $\pi/2$, $\pi$, and $\pi/2$ respectively~\cite{Kasevich1991}.  After the third pulse, the fraction of atoms in state $\ket{F=2,M_F=0}$ is given by

\begin{equation}
\mathcal{P}=\tfrac{1}{2}(1-\eta\cos(\varphi+\varphi_0)),
\label{eq:P}
\end{equation}
where \begin{equation}
\varphi=\frac{2\pi(f_1+f_2)}{c}\,T^2 \bold{n} \cdot \bold{a},
\label{eq:phase}
\end{equation}
and $\bold{n} \cdot \bold{a}$ is the acceleration of the atoms relative to the mirror, projected onto the mirror normal $\bold{n}$. The fringe visibility $\eta$ depends on the inhomogeneous width of the Raman transition, which in turn depends on the range of velocities selected. The additional angle $\varphi_0$ in Eq.~(\ref{eq:P}) is a phase shift applied to the last Raman pulse, which we switch between $0$ and $\pi$ on alternate shots of the experiment.  The difference in  $\mathcal{P}$ for these two phases is
\begin{equation}
\Delta\mathcal{P}=\eta\cos\varphi.
\label{eq:DeltaP}
\end{equation}
 Each interferometer sequence ends with a measurement of $\mathcal{P}$ using laser-induced fluorescence, and measurements alternate between having $\varphi_0=0$ and $\pi$ in order to determine $\Delta\mathcal{P}$, and hence $\varphi$. 
In total, $\bold{n} \cdot \bold{a}$ in Eq.~(\ref{eq:phase}) comprises three parts:
\begin{equation}
\bold{n} \cdot \bold{a}=\bold{n} \cdot (\bold{a}_{\text{ball}}+\bold{g}+\bold{a}_{\text{noise}}).
\label{eq:aparallel}
\end{equation}

\begin{figure*}
\centering 
\includegraphics[width=0.9\linewidth]{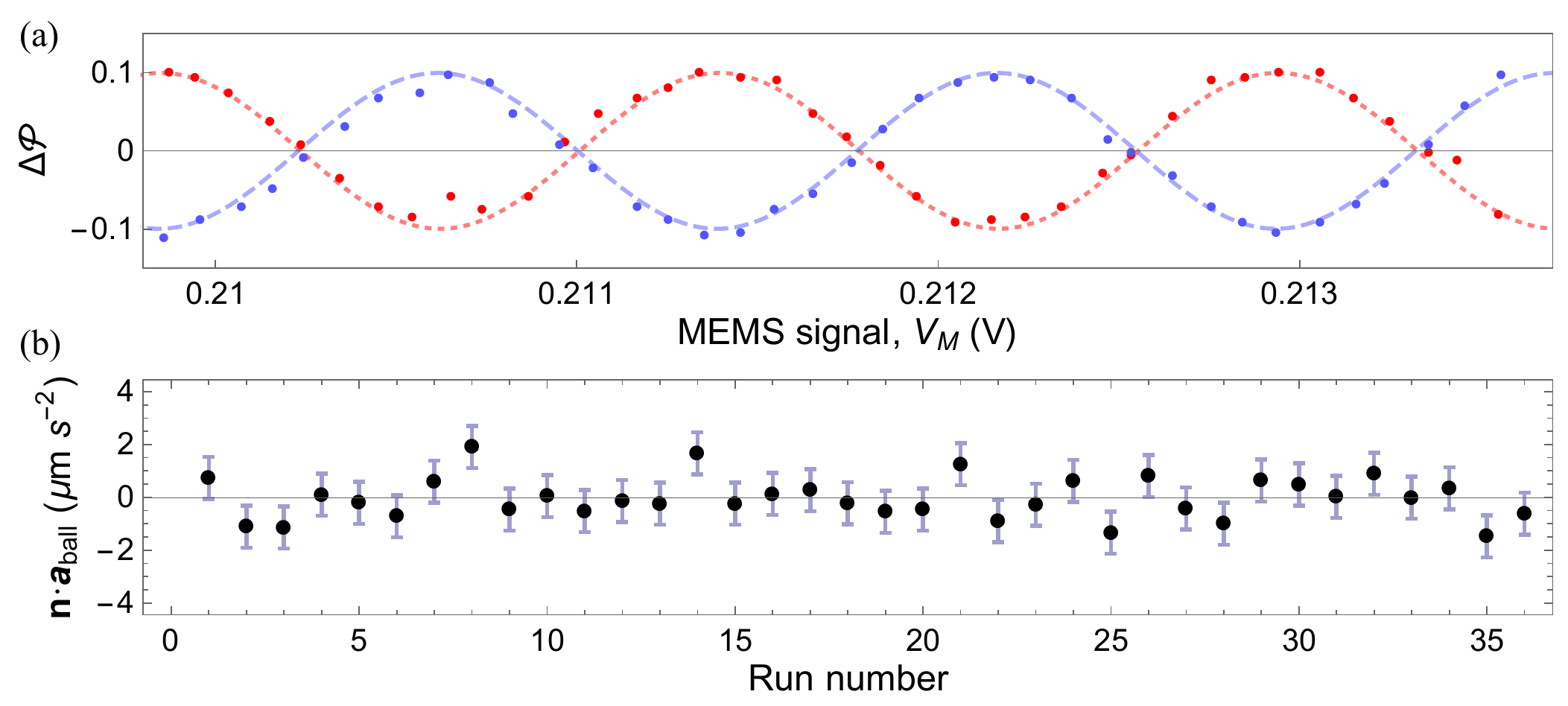}
\caption{
Experimental data. (a) Typical 12-hour run to measure $\Delta\mathcal{P}$ as a function of the voltage recorded by the FLEX accelerometer. This is scanned by tipping the table on which the atom accelerometer sits. Red points: ball in position 1. Here 5774 shots of the interferometer, gave 2887 measurements of $\Delta\mathcal{P}$, which have been averaged in bins of $100\,\mu$V width. Blue points: ball in position 2, data inverted for clarity.  Dashed lines: fits to Eq.~(\ref{eq:DeltaP}). (b) Our 36 independent measurements of $\bold{n} \cdot \bold{a}_{\text{ball}}$. Error bars indicate the standard deviation of these points.
\label{fig:data}}
\end{figure*}
The first term in parentheses, $\bold{a}_{\text{ball}}$, is the acceleration of the atoms toward the ball, including the normal Newtonian attraction and the anomalous acceleration $\bold{a}_{\phi}$ that we aim to measure. The second term is gravitational acceleration of the atoms toward the earth (and the rest of the environment). The last term is the acceleration noise of the mirror, due mainly to acoustic vibrations. In order to distinguish the first term from the other two, we mount a navigation grade FLEX accelerometer (Honeywell, QA750) on the back of the mirror, and plot the $\Delta\mathcal{P}$ fringes against the voltage $V_M$ registered by that accelerometer, as shown in Fig.~\ref{fig:data}(a). In this plot, the phase of the fringe pattern only senses changes in $\bold{n} \cdot \bold{a}_{\text{ball}}$, because the atomic and FLEX accelerometers both experience  the same  $\bold{n} \cdot (\bold{g}+\bold{a}_{\text{noise}})$ \footnote{The FLEX accelerometer is accurately oriented to measure along the normal to the mirror. Gradients of $\bold{g}$ produce a different $\bold{g}$ for the atoms and the FLEX accelerometer, but these are small enough to neglect.}, whereas the FLEX accelerometer, being far from the ball, does not register $\bold{n} \cdot \bold{a}_{\text{ball}}$. There can be a contribution to the phase of the fringe pattern from a zero offset (bias error) of the FLEX accelerometer, so we  alternate the position of  the source mass between positions 1 and 2,  which flips the sign of  $\bold{n} \cdot \bold{a}_{\text{ball}}$ but has no effect on the FLEX bias error \footnote{This does tilt the table, changing $\bold{n} \cdot \bold{g} $ by 125.4(1) nm/s$^2$, but that affects the atoms and the FLEX accelerometer equally so it does not appear as a shift of $\varphi$.}. The data points in Fig.~\ref{fig:data}(a) show a typical 12-hour set of fringes in $\Delta\mathcal{P}$ obtained by tilting the laser table (on which the experiment sits) in order to scan $\bold{n} \cdot \bold{g}$, and hence to scan $\varphi$. The red points are measured with the ball in position 1. The red dashed line is a fit to the function $\eta \cos{(\alpha V_{M}+\theta_1)}$, where $\alpha$ is the (measured) conversion from FLEX accelerometer voltage to atom interferometer phase, and $\theta_1$ is the phase offset due to $\bold{n} \cdot \bold{a}_{\text{ball}}$. Similarly the blue points are for the ball in position 2, with a fit to $\eta \cos{(\alpha V_{M}+\theta_2)}$. The difference between between $\theta_1$ and $\theta_2$ is so small that we plot the points for position 2 with their sign reversed for the sake of clarity.
 
Figure~\ref{fig:data}(b) shows the values for $\bold{n \cdot a}_{\text{ball}}$ obtained from 36 such measurements of $\theta_1-\theta_2$  taken during November 2017. On averaging these results we find that $\bold{n} \cdot \bold{a}_{\text{ball}}=-42\pm133$\,nm/s$^2$. By symmetry, we expect the average force to be along the line joining the centre of the atom cloud to the centre of the ball. Since this line lies at $48.8^{\circ} $ to the mirror normal $\bold{n}$, we infer that ${a}_{\text{ball}}=-64\pm201$\,nm/s$^2$, noting that for our switching sequence, the negative sign indicates a repulsive force. This raw result includes the Newtonian gravitational attraction, and also possible additions due to magnetic and electric field gradients, which we consider now and list in Table \ref{tab:systematics}.

\begin{table}[b]
\centering
\begin{tabular}{l|clll}
\hline\hline
 & Value (nm s$^{-2}$) \\
\hline
Measured $a_{\text{ball}}$ & $-64\pm201$ \\
Newtonian gravity & +7 \\
Magnetic field gradients &  $+6 \pm 5$  \\
Electric field gradients &  $<+2$ \\ \hline
Final value for $a_{\phi}$ & $-77\pm201$ \\
\hline\hline
\end{tabular}
\caption{Systematic corrections to $a_{\text{ball}}$, resulting in a value for $a_{\phi}$.}
\label{tab:systematics}
\end{table}

With a center-to-centre spacing of $26.8$\,mm, the Newtonian acceleration is $7.2$\,nm/s$^{2}$. Movement of the falling atoms through a magnetic field gradient can produce a false acceleration arising from the second-order Zeeman shift of the clock transition. We have surveyed the magnetic field of the ball \cite{SabulskyThesis} and conclude that the magnetic field gradient sourced by the ball contributes a false acceleration towards the ball of  $+6\pm5$\,nm/s$^2$. The electric analogue of this is negligible because the tensor polarisability, which determines the Stark shift of the clock transition, is exceedingly small. Instead, an electric field gradient can produce a real acceleration of the atoms toward the ball as a result of their scalar polarisability. The ball was grounded to the chamber, but it could have been electrically charged because aluminium naturally grows a surface layer of alumina that is $\sim4$\,nm thick and can support a potential difference of up to $\sim1$\,V before breaking down. Assuming that the surface is indeed charged to $\le 1$\,V, the acceleration of the atoms is $\le +1.5$\,nm/s$^2$. Collecting these results together in Table \ref{tab:systematics}, we obtain the final result 
\begin{equation}
a_{\phi} = -77\pm201\,\text{nm s$^{-2}$},
\end{equation}
where the uncertainty is almost entirely due to statistical noise. We conclude with 90\% confidence that  $a_{\phi}<+183$\,nm/s$^2$.

Using this upper limit on the strength of the non-Newtonian attraction, we now deduce limits on the parameters of chameleon and symmetron theories of modified gravity. Both theories have a scalar field action of the form (in the mostly-plus metric convention)
\begin{equation}
S_\phi = \int \mathrm{d}^4 x \sqrt{-g} \left(- \frac{1}{2} (\partial \phi)^2 - V(\phi) - A(\phi) \rho_\mathrm{m} \right)~,
\end{equation}
where $\rho_\mathrm{m}$ is the density of ordinary, non-relativistic matter, which is coupled directly to $\phi$ through the term $A(\phi)$. The equation of motion for $\phi$ follows from this action as
\begin{equation}
\vec \nabla^2 \phi = \frac{\mathrm{d} V}{\mathrm{d} \phi} + \frac{\mathrm{d} A}{\mathrm{d} \phi} \rho_\mathrm{m}~.
\label{eom}
\end{equation}

The scalar field $\phi$ mediates a fifth force between matter particles.  If a large object sources a scalar field  configuration $\phi(\vec x)$, then a small extended object in that field experiences an acceleration~\cite{Hui:2009kc}
\begin{equation}
\vec a_\phi = \lambda_\mathrm{a} \frac{\mathrm{d} A}{\mathrm{d} \phi} \vec \nabla \phi(\vec x)~.
\label{scalar-force}
\end{equation}
The screening factor $\lambda_\mathrm{a}$ approximately accounts for the backreaction of the small object on the field configuration $\phi(\vec x)$.  For sufficiently small and light objects, $\lambda_\mathrm{a} \to 1$ and the object is said to be unscreened.  On the other hand, large and dense objects decouple from the scalar field and satisfy $\lambda_\mathrm{a} \to 0$.  Here, the large and small objects are the metal ball and a $^{87}\text{Rb}$ atom, respectively.

\begin{figure*}
\centering 
\includegraphics[width=0.75\linewidth]{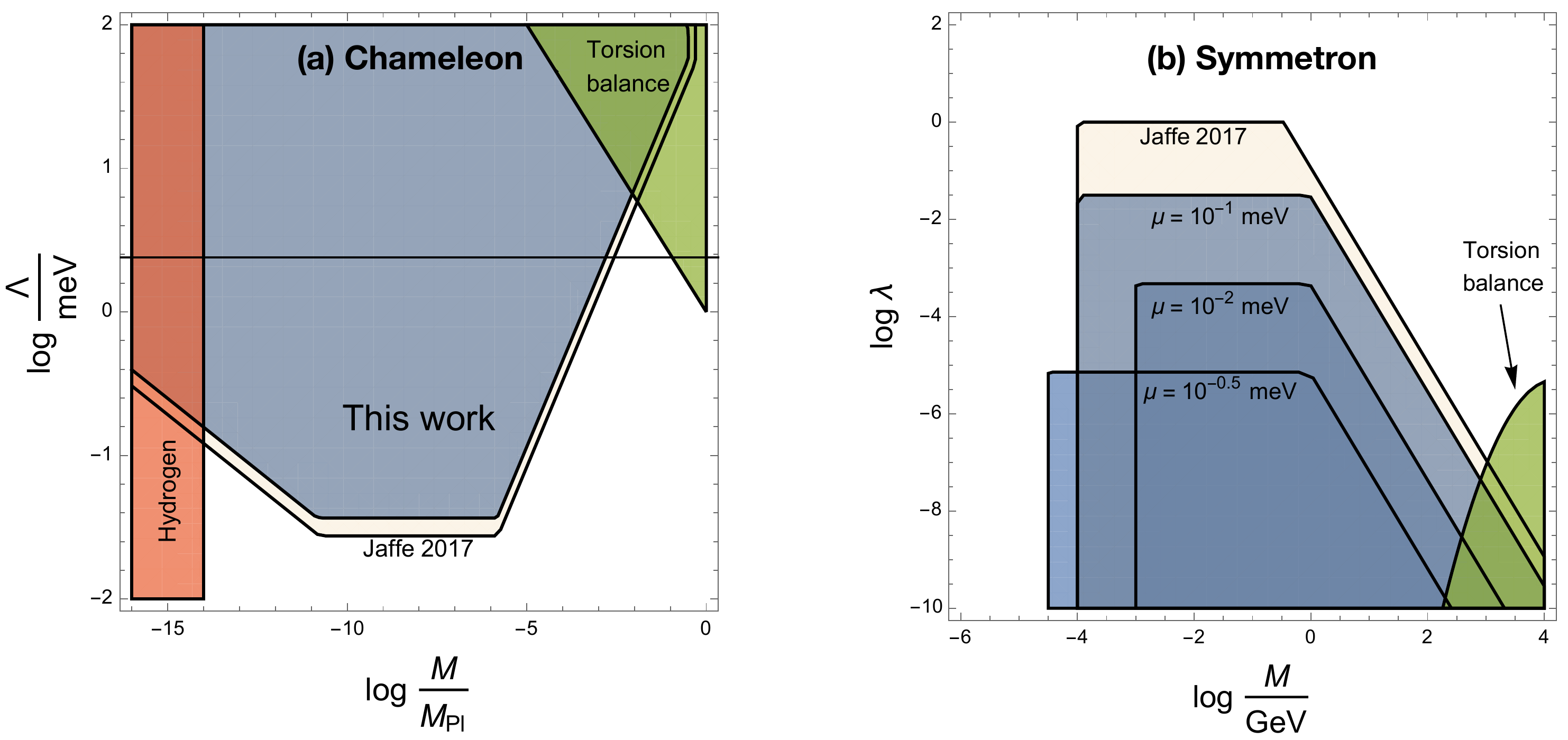}
\caption{Constraints on theory parameters for the chameleon and symmetron models, shaded regions are excluded.  The black line on (a) marks the dark energy scale $\Lambda = 2.4$ meV.   For the symmetron (b), only a range of approximately 1.5 orders of magnitude in $\mu$ is probed, as is typical of laboratory tests.  The Jaffe 2017 symmetron bounds apply for $10^{-1.5}$ meV $ < \mu \lesssim 10^{-1}$ meV, and the torsion balance constraints apply for $\mu \approx 10^{-1}$ meV.  A more complete listing of current bounds on both models may be found in \cite{Brax:2016wjk, Burrage:2017qrf,  Brax:2018zfb, Brax:2018grq}.\label{fig:exclusions} }
\end{figure*}

The prototypical example of a chameleon field has a self-interaction potential $V(\phi)$ and matter coupling $A(\phi)$
\begin{equation}
V(\phi) = \frac{\Lambda^5}{\phi}, \quad A(\phi) = \frac{\phi}{M}~,
\label{cham-model}
\end{equation}
and the screening factor for a spherical object of radius $R_\mathrm{obj}$ and density $\rho_\mathrm{obj}$ is approximately
\begin{equation}
\lambda_\mathrm{a, cham} \approx \min \left( \frac{3 M \phi_\mathrm{env}}{\rho_\mathrm{obj} R_\mathrm{obj}^2}, 1 \right)~,
\end{equation}
where $\phi_\mathrm{env}$ is the field value in the vicinity of the object.

Our aim is to solve Eq.~\eqref{eom} over the extent of the experiment, where $\rho_\mathrm{m}$ accounts for the source mass, the density of the gas in the vacuum chamber, and the vacuum chamber walls.  It is a non-linear partial differential equation, so an exact analytic solution for general $\rho_\mathrm{m}$ is impossible.  However, Eq.~\eqref{eom} is solvable for geometries with a high degree of symmetry, such as $\rho_\mathrm{m}$ corresponding to an infinite plate.  It may, of course, also be solved numerically. 

We adopt both the idealised-geometry and numerical approaches.  The distance from the atoms to the surface of the sphere is less than half the radius of the sphere, so we use a 1D parallel-plane approximation, as was done in \cite{Upadhye:2012qu,Upadhye:2012rc}.  In this approximation, the field is considered to be sourced by an infinite, very dense plate located at the nearest surface of the sphere.  The plate is assumed be so dense that $\phi \approx 0$ at the surface.   We also include an $O(1)$ geometrical fitting factor $\xi$, which is determined numerically.  This quantity accounts for deviation from the idealised parallel plate approximation, and is determined by the radius of the source mass and the geometry of the vacuum chamber walls.  This procedure avoids a computationally expensive search across many different parameter values, while still allowing us to enjoy the accuracy of the 3-dimensional numerical solutions.

The chameleon field is then approximated as \cite{Upadhye:2012qu,Burrage:2016lpu,Ivanov:2016rfs}
\begin{equation}
\phi_\mathrm{cham} = \xi_\mathrm{cham} (9 \Lambda^5 / 2)^{1/3} x^{2/3}~,
\label{cham-field}
\end{equation}
where $x = 0.775$ cm is the distance from the atoms to the nearest surface of the sphere.  It is easy to see that if $\xi_\mathrm{cham} = 1$, this solution satisfies the vacuum equation of motion in 1 spatial dimension.

We find the fitting factor $\xi_\mathrm{cham}$ by solving the full, 3-dimensional non-linear problem numerically.  We use a Gauss-Seidel finite-difference relaxation scheme, which accounts for the geometry of the vacuum chamber walls and source sphere.  The details of this code may be found in \cite{Elder:2016yxm, Jaffe2017}.

We assume that the walls and sphere are strongly screened, so we impose the boundary condition $\phi = 0$ at their surfaces.   We also assume the density of the gas in the vacuum chamber is negligible.  Both of these approximations are accurate within the main regions of interest.  Of course, the analysis presented here does not apply to regions of parameter space where these assumptions are not appropriate.  In those regions (for chameleon parameter $M \lesssim 10^{-10} M_\mathrm{Pl}$ and $M \gtrsim 10^{-0.5} M_\mathrm{Pl}$ with $M_\mathrm{Pl}$ being the Planck mass), we use an analysis identical to that of \cite{Burrage2015,Hamilton:2015zga} to place constraints.  Since those papers describe that method in detail, we do not include it here.

Comparing our numerical results to Eq.~\eqref{cham-field}, we find $\xi_\mathrm{cham} = 1.11$ across ten orders of magnitude of $\Lambda$, ranging from $\Lambda = 10^{-5}$ to $10^{+5}$ eV.
The independence of $\xi_\mathrm{cham}$ against $\Lambda$ is not a coincidence.  The chameleon equation of motion in vacuum admits the scaling symmetry
\begin{equation}
\phi \to a \phi~, \quad \Lambda \to a^{3/5} \Lambda~,
\end{equation}
so if $\phi$ is a solution to the equation of motion, then so is $a \phi$ but with the rescaled $\Lambda$.  Thus, the accuracy of Eq.~\eqref{cham-field} must be exactly the same  across all values of $\Lambda$.

Equation~\eqref{scalar-force} may now be used to compute the chameleon force, where we compute the screening factor for a rubidium-87 nucleus.  The constraints obtained from this expression are plotted in Fig.~\ref{fig:exclusions}(a).

The symmetron has a self-interaction potential and matter coupling
\begin{equation}
V(\phi) = -\frac{1}{2} \mu^2 \phi^2 + \frac{\lambda}{4} \phi^4~, \quad A(\phi) = \frac{\phi^2}{2 M^2}~,
\end{equation}
and the screening factor for a spherical object is~\cite{Hinterbichler2010}
\begin{equation}
\lambda_\mathrm{a, symm} \approx \min \left( \frac{M^2}{\rho_\mathrm{obj} R_\mathrm{obj}^2}, 1 \right)~.
\end{equation}
When the ambient matter density $\rho_\mathrm{m}$ is small, the field goes to the vacuum expectation value (VEV) $v = \mu / \sqrt{\lambda}$ at the minimum of its effective potential $V_\mathrm{eff} = V + A \rho$.   If the density is large, $\rho_\mathrm{m} > \mu^2 M^2$, the minimum of the effective potential is $\phi = 0$.  The scalar force Eq.~\eqref{scalar-force} is proportional to the local field value, so large ambient matter densities effectively shut off the scalar force.

Because the symmetron has a mass in vacuum, laboratory experiments typically only probe a relatively narrow range of $\mu$.   If $\mu$ is much larger than the inverse distance between the atoms and source mass, Yukawa suppression makes the symmetron force too short-ranged to appreciably affect the atoms.  On the other hand, if $\mu$ is smaller than the inverse length of the vacuum chamber, then it cannot reach the VEV anywhere in the vacuum chamber.  It is therefore energetically favourable for the field to remain at $ \phi = 0$ everywhere, shutting off the symmetron force \cite{Upadhye:2012rc}.  For this particular experimental setup, this window corresponds to $10^{-2} \mathrm{~meV} < \mu \lesssim 10^{-1} \mathrm{~meV}$, where the lower bound has been confirmed numerically.

Again, the approximate solution to the symmetron field is a product of a fitting factor $\xi_\mathrm{symm}$ and the 1D parallel plane solution \cite{Upadhye:2012rc,Brax:2017hna,Burrage:2018xta}
\begin{equation}
\phi_\mathrm{symm} = \xi_\mathrm{symm} (\mu / \sqrt{\lambda}) \tanh(\mu x / \sqrt{2})~.
\end{equation}

Like the chameleon, the symmetron equation of motion in vacuum admits the scaling symmetry
\begin{equation}
\phi \to a \phi~, \quad \lambda \to \frac{1}{a^2} \lambda~,
\end{equation}
which guarantees the stability of $\xi_\mathrm{symm}$ across different values of $\lambda$.  A similar argument does not apply for $\mu$, and therefore $\xi_\mathrm{symm}$ can fluctuate with $\mu$.  We have numerically solved the equation of motion for a number of values in the range $10^{-2}$ meV $ < \mu < 300$ meV, and found that $\xi_\mathrm{symm}$ is always between 1 and 1.5.  We therefore take the conservative approach and set $\xi_\mathrm{symm} = 1$.  The resulting constraints on the symmetron parameters are illustrated in Fig.~\ref{fig:exclusions}(b).

In conclusion, we have measured the acceleration of $^{87}$Rb atoms toward a test mass in a high-vacuum environment, using a significantly different method from previous experiments. Although the measurement should be sensitive to the force due to chameleon and symmetron fields, we find no evidence for the existence of such a force. Our result for the chameleon excludes almost the same parameter space as that of \cite{Jaffe2017}. In particular, if we take $\Lambda=2.4$\,meV we find that $\log(M/M_\mathrm{Pl})>-2.8$. The constraints we present for the symmetron parameters cover a wider range of $\mu$ than those of \cite{Jaffe2017}, but again the excluded region is similar. The parameter space covered by both experiments is now ruled out by two independent tests.

We acknowledge the engineering support of J. Dyne, S. Maine, G. Marinaro, and V. Gerulis. This work was supported by the UK research council EPSRC, Dstl, and the Royal Society. The European Commission supported D.O.S. during the majority of this work through a Marie Sk\l{}odowska Curie Early Stage Researcher program and the Action-Initial Training Network: Frontiers in Quantum Technology (FP7/2007-2013).  CB and BE are supported by a Leverhulme Trust Research Leadership Award.  EJC is supported by STFC Consolidated Grant No. ST/P000703/1.

\bibliography{references}
\end{document}